# Heating-induced strengthening or weakening of clays during slow to fast shearing at landslide stress levels

Marco Loche[1], Gianvito Scaringi[1,*]

[1] Institute of Hydrogeology, Engineering Geology and Applied Geophysics, Faculty of Science, Charles University, Albertov 6, 128 43 Prague, Czech Republic

* Corresponding author: gianvito.scaringi@natur.cuni.cz

## Abstract

Changes in temperature in landslide shear zones can derive from frictional heating but also groundwater flow or heat exchange with the atmosphere. This is especially relevant in shallow landslides under seasonal and long-term climatic forcing. How temperature controls the shear resistance in these landslides is poorly constrained. We explored the response of two "pure" clays (Ca-bentonite and kaolin) under landslide stress levels (50–150 kPa) and slow-to-rapid shearing (0.018–44.5 mm/min). We modified a ring-shear device to permit temperature control (20–50 °C). We observed important heating-induced strengthening in bentonite during slow shearing, but also weakening and again strengthening during faster shearing. Effects in kaolin were generally of opposite sign and lower magnitude compared to those in bentonite. Although more analyses are warranted in natural soils and in a lower temperature range, we argue that the thermal sensitivity of clays may matter not only in fast and large, but also in slow and shallow landslides in clay soils. However, case-by-case investigations should be performed to evaluate how thermal sensitivity combines with concurrent processes of atmosphere-soil interaction.

## Introduction

The velocity-weakening phenomenon has been explored extensively in large, long-runout landslides and seismic faults[1–5], with a variety of mechanisms having been put forward to explain their surprisingly low shear resistance[6–11]. Most of these mechanisms rely on the important frictional heating within the shear zone, where temperature sufficient for steam generation (in water-rich conditions) or even mineral alteration (with carbon dioxide generation or mineral melting) can be attained. In slow-moving landslides, the contribution of frictional heating is negligible as heat conduction and convection efficiently prevent temperature from rising; therefore, temperature-dependent processes are typically not expected. However, other causes of variation in temperature exist – both endogenous and related to changing boundary conditions. The former can entail, for instance, endo- or exothermic chemical and biological processes[12]; as for the latter, both direct and coupled flows can play a role: for instance, the flow of groundwater not only defines pore water pressures but also alters the thermal energy balance.

Notably, although a variety of processes exist, that can potentially alter temperature patterns within the ground, with consequences in terms of altered hydraulic and mechanical properties, little has been done to explicitly account for them in landslides and in most geotechnical problems[10]. One of the research directions bringing awareness to the need of explicitly accounting for temperature in geomechanics is the shear rate-dependency of friction, which is typically evident under very large shear rates in the form of frictional weakening. However, also under comparatively low shear rates, different rate effects have been recognized, which were termed "positive" (rate strengthening), "negative" (rate weakening), and "neutral" in a classic paper by Tika et al.[13]. These effects seems to be associated with changes in the mode of shearing, with turbulent shearing producing a neutral or negative rate effect, laminar shearing being typical of a rate-strengthening behavior, and an intermediate, transitional mode, in which a negative rate effect prevails. More recently, Scaringi et al.[5,14] collated a large number of experimental results in the full range of explorable shear displacement rates, corresponding to extremely slow to extremely rapid landslide movements. The authors pointed out further complexity (non-monotonicity) in the rate-dependent response: soils may, for instance, exhibit weakening followed by strengthening at low to intermediate shear rates, and finally dramatic weakening at large rates (>$10^{-2}$ m/s), but only under large confining stresses (>0.5–1 MPa), related to frictional heating and pressurization[5,14–16]. Results in a range of confining stresses (typical of shallow to deep-seated landslides), shear rates (typical of extremely slow to rapid landslides), experimental devices, and soil compositions are often complex and difficult to interpret comprehensively[14,17–19]; nonetheless, laminar shearing and the consequent shear rate-strengthening seem the most common behaviors in clay-rich soils, regardless of their specific mineralogy.

Temperature adds a further layer of complexity to the interpretation of these and other geotechnical experiments, as clearly pointed out in the seminal work by Mitchell[20]. In a recent review[10], it was highlighted how a variegated literature exists on temperature effects in geomaterials which, however, lacks systematicity and clear field evidence. One exception is the work of Shibasaki et al.[21,22], in which the behavior of a number of slow-moving landslides accelerating during late autumn without a hydro-mechanical cause could convincingly be explained by a cooling-related shear-weakening of the clay-rich soil, as demonstrated by temperature- and shear rate-controlled experiments. Their work, however, remains an isolated case, surrounded by other publications in which thermal effects keep being considered negligible and remain unexplored[23–28]. Additionally, a recent review[29] pointed out that most published results on fine-grained soils consider the friction angle independent of temperature[27,28,30–34]. A similar conclusion can be found in other states of the art[35–37]. On the other hand, it stands out that a specific geotechnical literature, that focuses on the behavior of engineered clay barriers as well as clay host formations in deep geological radioactive waste repositories, is well aware of the key role of thermal effects in controlling the hydro-mechanical response of the geomaterial in question[38–41].

We argue that, in large-deformation problems, such as in landslide behavior, the role of temperature should be quantified systematically. What is more, this role and the consequent effects on the landslide fate should be evaluated together with that of shear-rate effects. In this work, following the path tracked by Shibasaki et al.[21,22], we report on results of ring-shear tests on water-saturated samples of two very different clays (a Ca-bentonite and a kaolin), performed under various stress levels and shear rates, typical of slow to rapid landslides. We evaluated the residual shear strength in usual condition (20 °C), then increased the temperature up to ~50 °C by transferring heat into the device's water bath.

**Results and discussion**

**Figure 1** shows the experimental results in terms of residual friction coefficient (shear strength normalized by the normal stress). The blue lines represent changes in friction, which are compared to the changes in temperature in the device water bath (red lines). What first stands out is that the behavior of bentonite (**Figure 1a**) differs significantly from that of kaolin (**Figure 1b**); furthermore, the magnitude and sign of the temperature effect changes with the shearing rate. At the lowest rate (0.018 mm/min) the bentonite shows a positive temperature effect (strengthening upon warming), consistently with its high smectite content, as already observed by Shibasaki et al.[21,22]. The effect seems completely reversible; however, upon closer look, it can be seen that the strength increase is slightly delayed (~1 mm corresponds to ~1 h of testing at this shear rate) and, at the end of cooling, the available resistance is slightly lower than that available before heating. A very slow recovery then occurs over a distance of ~10 mm (~10 h). This seems compatible with a mechanism of heating-induced consolidation[42] of the shear-zone material which, owing to the large shearing strains, acquires a normally-consolidated structure irrespective of the pre-shearing stress history. This mechanism is accompanied by generation of excess pore water pressure that, owing to the very low hydraulic conductivity of the bentonite, is still dissipating after the completion of the heating-cooling cycle. The subsequent cooling-induced rebound has been shown in the literature to only cause a small volumetric recovery[20]. Furthermore, this pore water pressure excess may reduce, to some extent, the magnitude of the temperature effect.

As for the material far from the shear zone, its large overconsolidation makes it susceptible to heating-induced swelling and subsequent contraction upon cooling, with an opposite effect on pore water pressure compared to that within the shear zone. The hydraulic interaction between these two zones of the sample can only be elucidated by directly measuring the pore water pressures (e.g., on the top – shear zone – and bottom – far from the shear zone – of the samples) and carefully controlling normal displacements in absence of soil extrusion, which would require a more capable experimental device. Nonetheless, it can be presumed that the large pressure gradient across the sample can only facilitate the dissipation of excess pressure, thus the magnitude of the peak in strength due to heating seems reasonable. During the period at constant high temperature (sufficiently long in the tests under 100 and 150 kPa), it is also possible to observe a slow decrease in strength over time, seemingly towards a final value (unreached) that should nevertheless be quite higher than the pre-heating residual shear strength. Provided that this kind of behavior (softening) cannot be attributed to the aforementioned pore water pressure dissipation, which would cause an opposite phenomenon, it needs to be attributed to a mechanism acting under constant effective stress, such as the post-peak softening typically observed in overconsolidated clay. This has two possible roots: a rearrangement of matter caused by shearing in the heated (and hence apparently overconsolidated) material, and/or the involvement of further material, which is overconsolidated, into the (thinned-by-heating) shear zone. Extension of the testing campaign to normally-consolidated samples could clarify this matter. However, testing complications would arise, owing to the much larger compressibility of the samples and the consequent possible excessive thinning (resulting in important lateral friction) and extrusion.

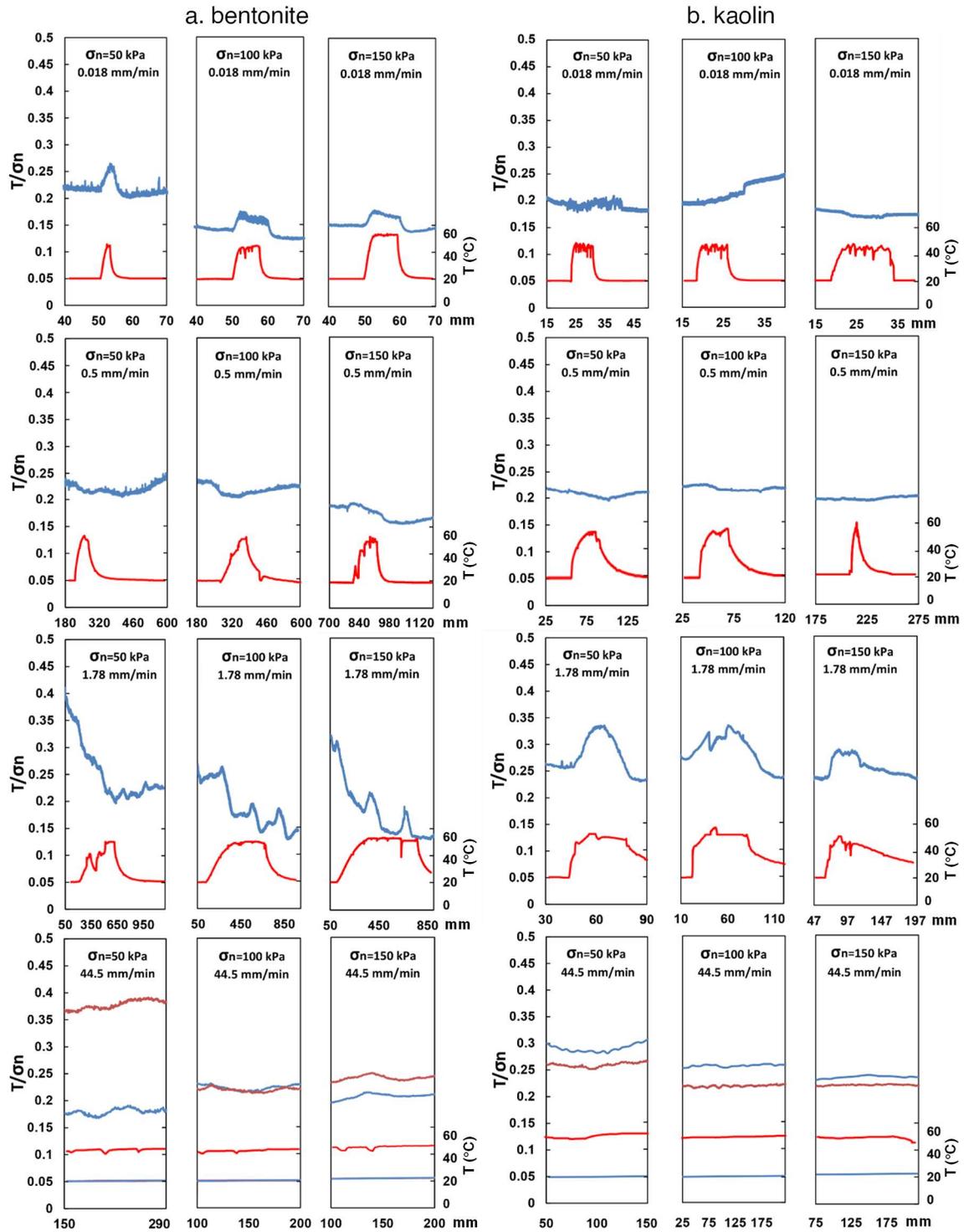

**Figure 1.** Results of the ring-shear tests on a) bentonite and b) kaolin. The tests were performed under normal stresses of 50, 100, and 150 kPa, corresponding to overconsolidation ratios of 12, 6, and 4 (lost in the shear zone upon shearing), and equivalent linear shear rates of 0.018, 0.5, 1.78, and 44.5 mm/min. Note that, for the highest rate, two distinct samples were used to test the effect of temperature.

Moving to the slow (0.018 mm/min) test performed on kaolin, a different temperature-dependent behavior can be observed, with some negative temperature effect (weakening by warming) only observed under a confining stress of 150 kPa, for which stress values are large enough to obtain low-noise results. Noise is apparent under the lowest confining stress, which seems to completely mask any temperature effect. Similarly, no temperature effect can be discerned in the test under 100 kPa, during which a clear residual condition possibly had not yet been attained. As for the tests performed at a higher rate (0.5 mm/min) some negative temperature effect can be seen in bentonite and, even less accentuated, in kaolin. Note again that the recovery to the pre-heating condition occurs well after the completion of the heating-cooling cycle (note the different horizontal scale in the figure). With the available information, it is not possible to discern whether the small strength loss can be caused by a pore-water pressure excess or signals a true heating-induced shear weakening, which could be consistent with a change in shear mechanism from laminar to turbulent[22].

Under an even larger shearing rate (1.78 mm/min), for which turbulent shearing should be the norm under the explored confining stresses[13,19], we observed a rather disturbed shearing in bentonite, possibly due to the development of irregularities on the shear surface[5]. Some irregularities also are observed during the test on kaolin. In both materials, however, important changes in shear resistance are observed: weakening in bentonite and strengthening in kaolin upon heating, with an effect that is surely reversible in the latter. In bentonite, it is possible that the heating-induced compaction in the shear zone may have favored some regularization, with better alignment of the particles, conducive to a permanent weakening. However, further experiments are warranted to exclude extemporaneous behaviors. Finally, under the largest shearing rate (44.5 mm/min), at which two distinct tests were conducted for each material (one at room temperature and one at elevated temperature), positive and negative temperature effects are observed in bentonite and kaolin, respectively, similar to those observed at the lowest shearing rate (0.018 mm/min). This differs from the pattern reported in the literature[22] and would suggest that, even within the realm of turbulent shearing, temperature effects may still change with the shearing rate in a given clay. On the other hand, both the shear rate and temperature effects are difficult to interpret under fast shearing in saturated clay samples, as changes in pore water pressures are occurring together with important disruptions in the shear zone. These tests would be best performed in larger ring-shear devices with pore water pressure control (e.g., undrained shearing), that should be modified ad hoc to allow for careful temperature control. To the authors' knowledge, such devices have not yet been realized.

A summary of the results is shown in **Figure 2**. In particular, **Figure 2a** and **2b** show the failure envelopes of bentonite obtained at 0.018 and 0.5 mm/min, for which well-defined trends were evaluated. At the lowest rate, an increase in friction coefficient by about 13% is evaluated (from ~0.15 to ~0.17). At the higher rate, only a slight decrease in friction coefficient remains (-2%, from ~0.22 to ~0.21). The variability of the temperature effect is well apparent in **Figure 2c** and **2d**, which exhibit important scattering under a given shearing rate (with varying normal stress) as well as changes in behavior across the rates. Notably, thermal effects seem more conspicuous in bentonite than in kaolin; furthermore, they tend to be of opposite sign at any rate, and this is consistent with the inversion of thermal sensitivity[22] as a function of the soil's smectite content (from negative to positive temperature effect as the proportion of smectite increase). It is worth noting, however, that both our tests and those reported in the literature used Ca- or Ca-Mg-bentonite, whereas the behavior of K-bentonite and the much more active Na-bentonite remain unexplored. Similarly, illitic and illitic-smectitic soils have not been investigated and should be the object of future systematic work. Future work should also focus on exploring rate and temperature effects in a lower temperature range (approaching 0 °C), more common in temperate and cold climates, and in a different experimental setup, in which stresses rather than displacement rates are imposed (so-called shear creep tests, which more closely resemble actual field conditions in slow-moving landslides). Such tests have been performed in the past to explore shear rate effects in the form of relaxation tests[5] or changes in available shear strength due to chemical forcing[43–45]. Thus far, no such tests have been performed under variable temperature.

In summary, we showed results of shear rate- and temperature-controlled ring-shear experiments on a very active and a non-active clay with the purpose of evaluating a possible mineralogy-dependent coupling between the two imposed conditions. We observed significant strengthening upon heating in Ca-bentonite under slow shearing (0.018 mm/min) and slight weakening in kaolin. The effects were, however, of comparatively small magnitude, in the order of 0.1–0.5%/°C of shear strength change, while effects up to 1.5%/°C were expected in the bentonite on the basis of its smectite content[22]. We argued that pore water pressure excess arising in the shear zone owing to heating-induced consolidation may explain an apparent reduction in the thermal sensitivity of the residual shear strength, and the mechanism is consistent with the observed delayed response to natural cooling after the high-temperature phase. Different effects were evaluated in bentonite and kaolin upon heating during faster shearing (0.5, 1.78, and 44.5 mm/min). Notably, we observed a non-monotonic response of opposite sign in the two material, with a negative temperature effect in bentonite and a positive effect in kaolin, turning back to positive and negative effects, respectively, at the highest shearing rate (44.5 mm/min). Non-monotonic responses as a function of the shearing rate have been reported previously[5]; however, owing to the large uncertainty deriving from turbulent shearing in the used apparatus and the unaccounted pore water pressure changes, further experiments are warranted.

Overall, the presence of thermal sensitivity in clays and the large differences in both temperature and shear-rate effects depending on mineralogy suggest that the characterization of clay soils under shearing could be improved by explicitly accounting for temperature and/or exploring the temperature-dependency of strength parameters. In landslide studies, even in slow-moving landslides and especially in shallow movements, accounting for thermo-mechanical coupling in modelling could help better predict the (re)activation and runout of unstable bodies.

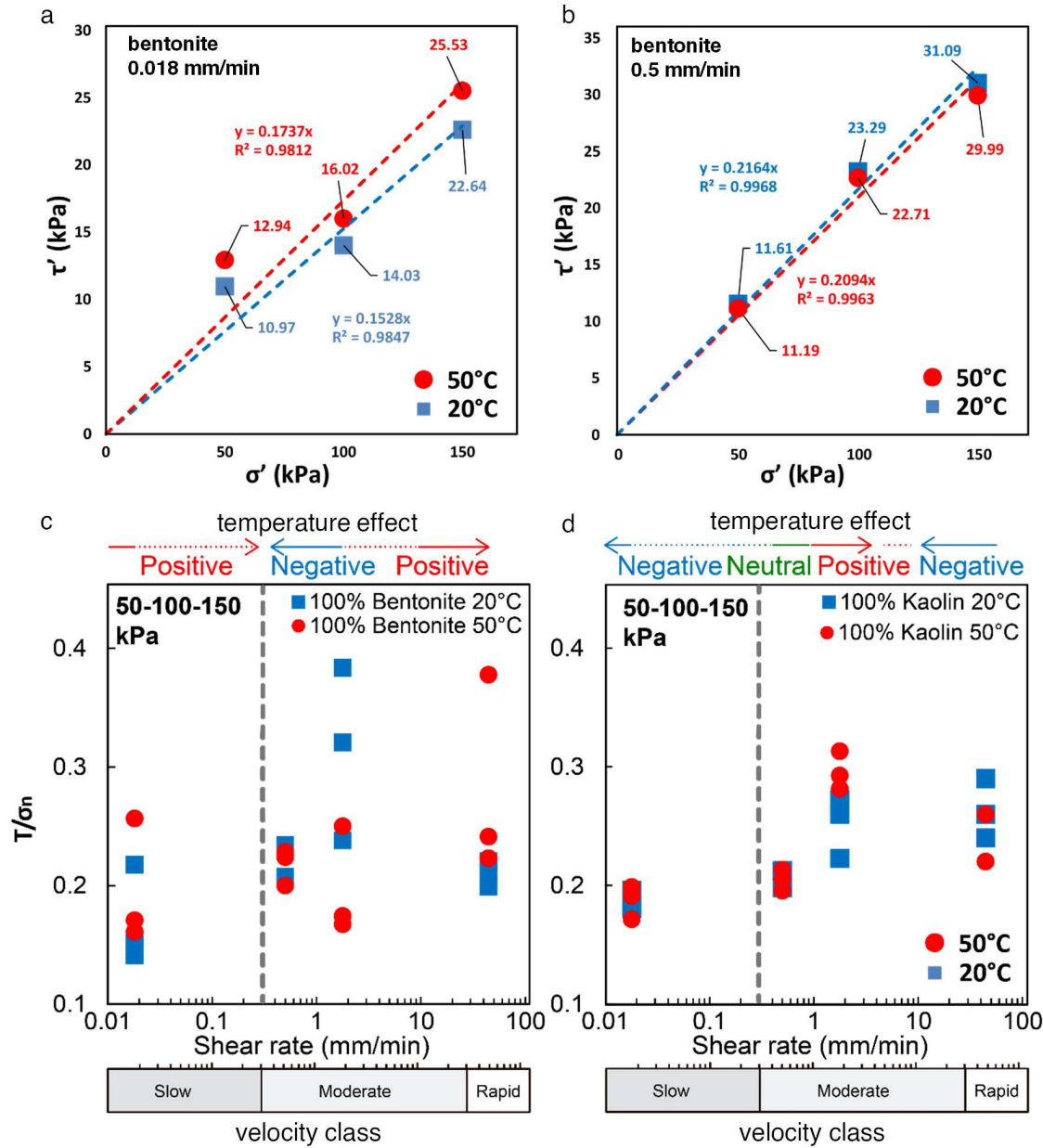

**Figure 2. Summary of temperature effects.** a) and b) show the residual shear strength envelope for bentonite at 0.018 and 0.5 mm/min, respectively: an increase in temperature from 20 °C to 50 °C caused an increase in friction coefficient from ~0.15 to ~0.17 (+13%) at 0.018 mm/min but a decrease from ~0.22 to ~0.21 (-3%) at 0.5 mm/min. A summary of all test results is provided in c) and d) for bentonite and kaolin, respectively.

## Methods

We employed commercially available bentonite and kaolin. In particular, we chose the Czech B75 bentonite from the Černý vrch deposit, having ~85 wt% of Ca-Mg montmorillonite, liquid limit $w_L = 217\%$, plastic limit $w_P = 51\%$, and specific gravity of the solids $G_s = 2.87$. In the sieving-sedimentation analysis, 61 wt% was <2 μm, with a further 33 wt% <60 μm, qualifying the material as a CH clay [46] with activity $A = 2.7$. Further characterization can be found in the literature[47,48]. The kaolin was a refined Malaysian FMC kaolin from Perak, with ~46 wt% kaolinite and ~48% wt% muscovite, $w_L = 63.2\%$, $w_P = 40.1\%$, and $G_s = 2.65$. In the sieving-sedimentation analysis, 28 wt% was <2 μm, with a further 70% <60 μm, defining it as an MH silt[46] with $A = 0.82$.

We performed the tests in a conventional ring-shear apparatus[49,50] equipped with a temperature-change device allowing circulation of water in a closed circuit between an external temperature-controlled bath and the shear-box bath. The device accommodates a 5 mm thick annular sample, sandwiched between roughened brass porous platens, to ensure drainage and avoid interface shearing. We minimized the lateral friction by ensuring post-consolidation sample thicknesses around 4 mm[51]. We exploited the available range of rotational rates, producing equivalent linear rates of 0.018–44.5 mm/min which are associated with slow (m/yr) to rapid (mm/s) landslide movements[52,53]. We prepared the reconstituted samples following Burland[54] and consolidated them stepwise while ensuring the dissipation of pore pressure excess by monitoring the consolidation curve[55]. The sample was consolidated until 600 kPa and subsequently unloaded to reach the required normal stress (50-100-150 kPa). Under each stress level, the shear rate was increased stepwise, and an overconsolidated material was produced. With this procedure, a structural anisotropy can be produced, in which the clay particles are already preferentially oriented in the direction of shearing (orthogonal to the direction of consolidation), thereby facilitating the generation of a smooth shear surface. To keep the required thickness and ensure smoothness of the shear surface for as long as possible during testing, we adopted a single-stage rather than a multi-stage procedure[56], which has been proven to provide accurate results, although at the expenses of the total testing time. The samples were kept in a bath of deionized water (<1 μS/cm) at room temperature (20±1 °C) and atmospheric pressure throughout the first part of the test. Once the residual stage was reached, temperature was gradually increased by activating a resistance in the thermal bath, to reach 50±5°C in the sample. The temperature was held constant while producing a sufficient shear displacement. Finally, the resistance was turned off, and natural cooling down to room temperature occurred while shearing was continued. **Figure 3** shows the general testing scheme, which is similar to that proposed in the literature[22], as well as the geometry of the testing device and its main parts. Note, however, that for tests under 44.5 mm/min, sample degradation occurs too fast, thus a heating-cooling cycle is not feasible. In this case, two distinct tests were performed: in one test, fast shearing was performed at room temperature; in the other test, fast shearing was performed after bringing the sample to the desired high temperature.

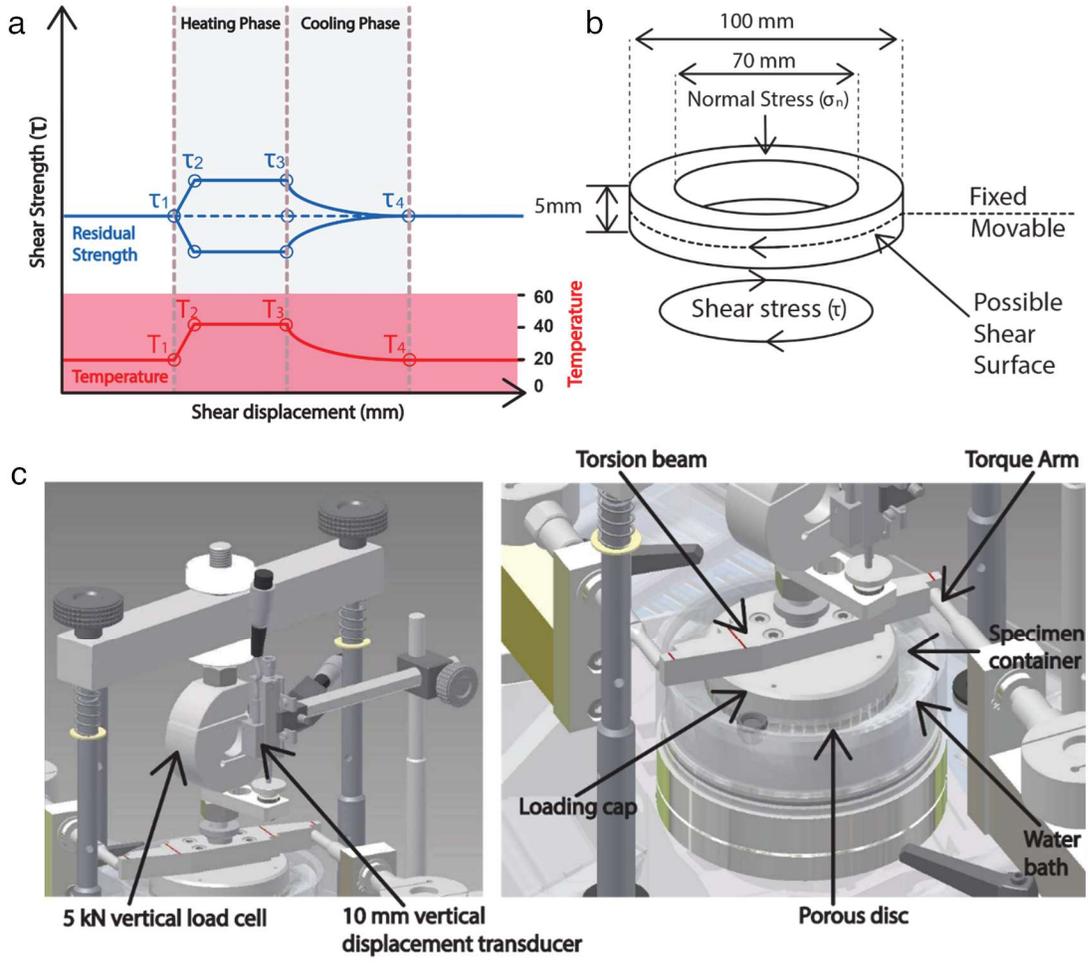

**Figure 3. Testing setup**. a) The shear experiments entail the attainment of the residual strength at room temperature under given normal stress and shear rate, followed by heating, constant high temperature, and cooling while keeping the shear rate unchanged [22]; b) geometry of the sample; c) components of the device.

## Acknowledgements

The authors acknowledge financial support by the Charles University Grant Agency (GAUK; Project Number 337121), the Grant Agency of the Czech Republic (GAČR; Project Number 20-28853Y), and the Fund for international mobility of researchers at Charles University (MSCA-IF IV; Project Number CZ.02.2.69/0.0/0.0/20_079/0017987).

## Data availability

All the elaborated data are contained in the paper. Raw data are available from the authors upon reasonable request.

## Author contributions

ML and GS planned the experiments, which were performed and interpreted by ML. ML prepared the figures and the original manuscript draft. GS further interpreted the results, edited the figures, and finalized the draft.

## Competing interests

The authors declare no competing interests.